\documentclass[aps,apl,twocolumn, notitlepage, superscriptaddress]{revtex4-1}
\usepackage[caption=false]{subfig}
\usepackage{color}
\usepackage{graphicx}
\usepackage{natbib}
\usepackage{amsmath}
\usepackage{ulem}
\usepackage{braket}
\usepackage[utf8]{inputenc}
\usepackage{url}
\usepackage{hyperref}
\usepackage{amssymb}
\usepackage{bbm}

\begin{document}
\definecolor{darkgreen}{rgb}{0,0.5,0}
\newcommand{\comment}[1]{\textcolor{darkgreen}{#1}}
\newcommand{\commentImp}[1]{\textcolor{red}{#1}}
\newcommand{\todo}[1]{\textcolor{blue}{ToDo: {#1}}}
\newcommand{\h}[1]{\hat{#1}}
\newcommand{\sinc}{\mathrm{sinc}}

\title{Limits of the time-multiplexed photon-counting method}
\author{Regina Kruse}
\email{regina.kruse@upb.de}
\affiliation{Applied Physics, University of Paderborn, Warburger Straße 100, 33098 Paderborn, Germany}
\author{Johannes Tiedau}
\author{Tim J. Bartley}
\author{Sonja Barkhofen}
\author{Christine Silberhorn}

\date{\today}

\begin{abstract}
The progress in building large quantum states and networks requires sophisticated detection techniques to verify the desired operation. To achieve this aim, a cost- and resource-efficient detection method is the time multiplexing of photonic states. This design is assumed to be efficiently scalable; however, it is restricted by inevitable losses and limited detection efficiencies. Here, we investigate the scalability of time-multiplexed detectors under the effects of fiber dispersion and losses. We use the distinguishability of Fock states up to $n=20$ after passing the time-multiplexed detector as our figure of merit and find that, for realistic setup efficiencies of $\eta=0.85$, the optimal size for time-multiplexed detectors is 256 bins.
\end{abstract}

\maketitle
\section{Introduction}
In recent years, the progress in both quantum source \cite{mosley_heralded_2008, eckstein_highly_2011, harder_optimized_2013, collins_integrated_2013, kumar_controlling_2014, bruno_pulsed_2014, jin_efficient_2015, fortsch_highly_2015, zielnicki_engineering_2015, harder_single-mode_2016} and detector \cite{lita_counting_2008,marsili_detecting_2013} engineering has placed the implementation of large photonic network structures into reach. However, in order to verify the intended operation of the networks, reliable multiphoton measurements are necessary to both check the state generation and measure the output distribution.

One method to measure the photon-number properties of a quantum state is true photon-number-resolving detectors, such as superconducting transition-edge detectors (TESs) \cite{lita_counting_2008}. However, while they offer intrinsic photon-number resolution, they are also resource demanding and require very low operating temperatures in the mK regime to work properly. Furthermore, their photon-number resolution is limited to a few tens of photons as the superconducting circuit breaks off at some critical energy \cite{fukuda_titanium-based_2011, gerrits_extending_2012, humphreys_tomography_2015}. Since intrinsic photon number resolving detectors have been around for only a short while and require a lot of resources, \textit{quasi}photon-number-resolving detectors have been proposed as a resource-efficient and cheap alternative. They use conventional on-off detectors (e.g., avalanche photodiodes) with either a spatial or temporal multiplexing scheme \cite{achilles_fiber-assisted_2003, rehacek_multiple-photon_2003, fitch_photon-number_2003, achilles_photon-number-resolving_2004, waks_direct_2004, jiang_photon-number-resolving_2007, heilmann_harnessing_2016}. In general, time multiplexing can be seen as the more resource-efficient technique, as the scheme allows us to use the same detectors again and again at the cost of increased measurement time, instead of using each detector only once.

\begin{figure}
\includegraphics[width=.95\columnwidth]{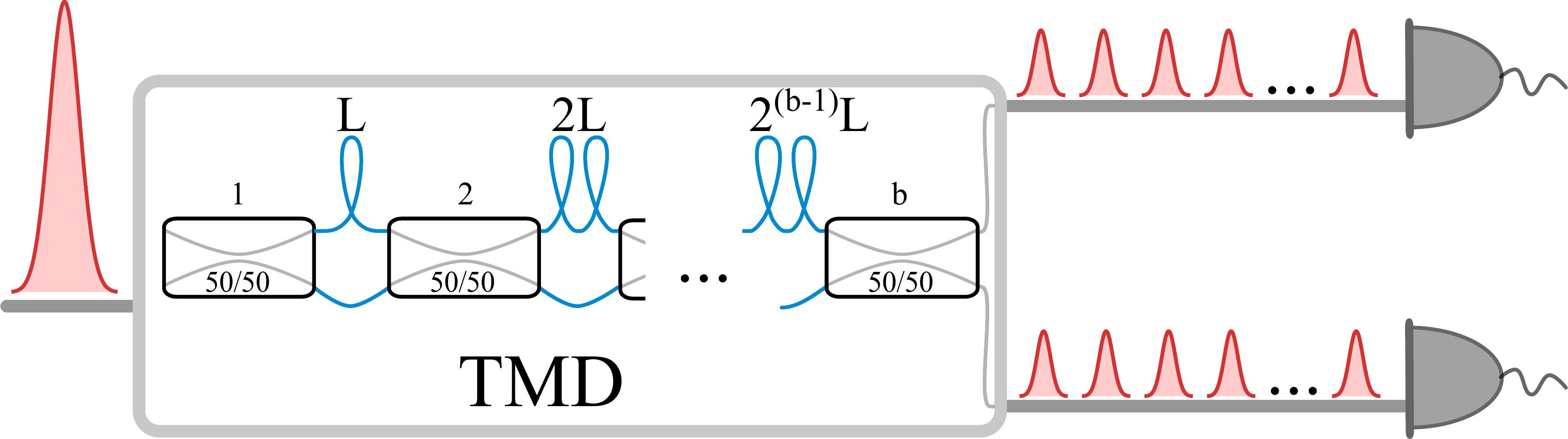}
\caption{(colour online) Schematic of a time-multiplexed detection network.}
\label{fig:loopy}
\end{figure}

The time-multiplexing network \cite{achilles_fiber-assisted_2003, rehacek_multiple-photon_2003, achilles_photon-number-resolving_2004} consists of several (generally fiber-integrated) beam splitters that are connected by different fiber lengths (see fig. \ref{fig:loopy}). While passing the network, one part of the input pulse is partially delayed in each beam splitter stage, such that pulses with $2^b$ different timings (time bins) arrive at the detector.

This method is scalable in principle, as the connection of fiber-integrated components does not pose an unsurmountable problem. In this context, Sperling \textit{et al.} \cite{sperling_true_2012} have considered the minimum size of a multiplexed detector to distinguish between different states in loss-free implementations. However, up to this point no investigation of the scalability has considered realistic, i.e., lossy and dispersive, fiber-integrated components and their effect on the detected photon number statistics.

In this paper, we answer how scalable the time-multiplexing photon-counting detection method really is. We approach this question from an experimental point of view by simulating the photon-number statistics after passing the time-multiplexing detector (TMD) and compare the quality of the TMD measurement by attempting to differentiate and reconstruct different Fock state inputs. As a figure of merit, we use the overlap of the simulated photon statistics after the TMD between neighboring Fock states and find the optimal size of the network.

This paper is structured as follows: in Sec. II, we give a preliminary limitation of the network size, as determined by the fiber dispersion and the input pulse length. Section III contains a discussion of the optimal network size considering state-of-the-art fiber-integrated components. We discuss the effects of distributing photon-number states onto a final set of time bins and losses in the beam-splitter network. Taking these two effects into account, we arrive at a practical limitation of the network size which gives the most reliable experimental data. Finally, in Sec. IV, we summarize our findings and conclude this paper.

\section{Limitation via Dispersion}
In this section, we consider the geometrical limitations of the TMD detection system, as given by the dispersion of the utilized fibers. To do this, we assume no-loss fibers and detectors with unit detection efficiency and vanishing dead time. In this (admittedly unrealistic) scenario the maximum number of available bins is reached when neighboring pulses start to overlap significantly at the output. This means that we determine this number by calculating how many output pulses fit in the time between consecutive experiments. Expressing the duration of a single shot in terms of the repetition rate of the experiment $R_\mathrm{rep}=(\Delta\tau_\mathrm{exp})^{-1}$, we define the maximal number of time bins as
\begin{equation}
N_\mathrm{max,\, disp}=\frac{\Delta\tau_\mathrm{exp}}{\Delta\tau_\mathrm{disp}}=(R_\mathrm{rep}\Delta\tau_\mathrm{disp})^{-1}\, ,
\label{eq:bins}
\end{equation}
where $\Delta\tau_\mathrm{disp}$ is the pulse width (FWHM) at the output after experiencing fiber dispersion. This effect is taken into account by the group-velocity dispersion \cite{diels_ultrashort_2006}
\begin{equation}
\Delta \tau_{\mathrm{disp}}=\Delta \tau_{\mathrm{in}}\sqrt{1+\left(\frac{4\,\mathrm{ln}(2)}{\Delta\tau_\mathrm{in}^2} \frac{\lambda^2}{2\pi c} D_\lambda L_\mathrm{exp}\right)^2}\, .
\label{eq:dispersion}
\end{equation}
Here, $\Delta\tau_\mathrm{in}$ (ps) is the FWHM pulse length of the input photons, $\lambda=1550\,$nm is the operating wavelength, $D_\lambda$ is the dispersion coefficient and $\Delta\tau_\mathrm{disp}$ (ps) is the output pulse length after passing a fiber of length $L_\mathrm{exp}=c \Delta\tau_\mathrm{exp}$ (km), with $c$ being the speed of light in the fiber. The dispersion coefficient $D_\lambda$ for standard SMF28 fibers is specified as $D_\lambda\leq18.0\frac{\mathrm{ps}}{\mathrm{nm}\cdot\mathrm{km}}$ \cite{corning_optical_fibre_optical_2007} at $1550\,$nm.
\begin{figure}
\includegraphics[width=1.\columnwidth]{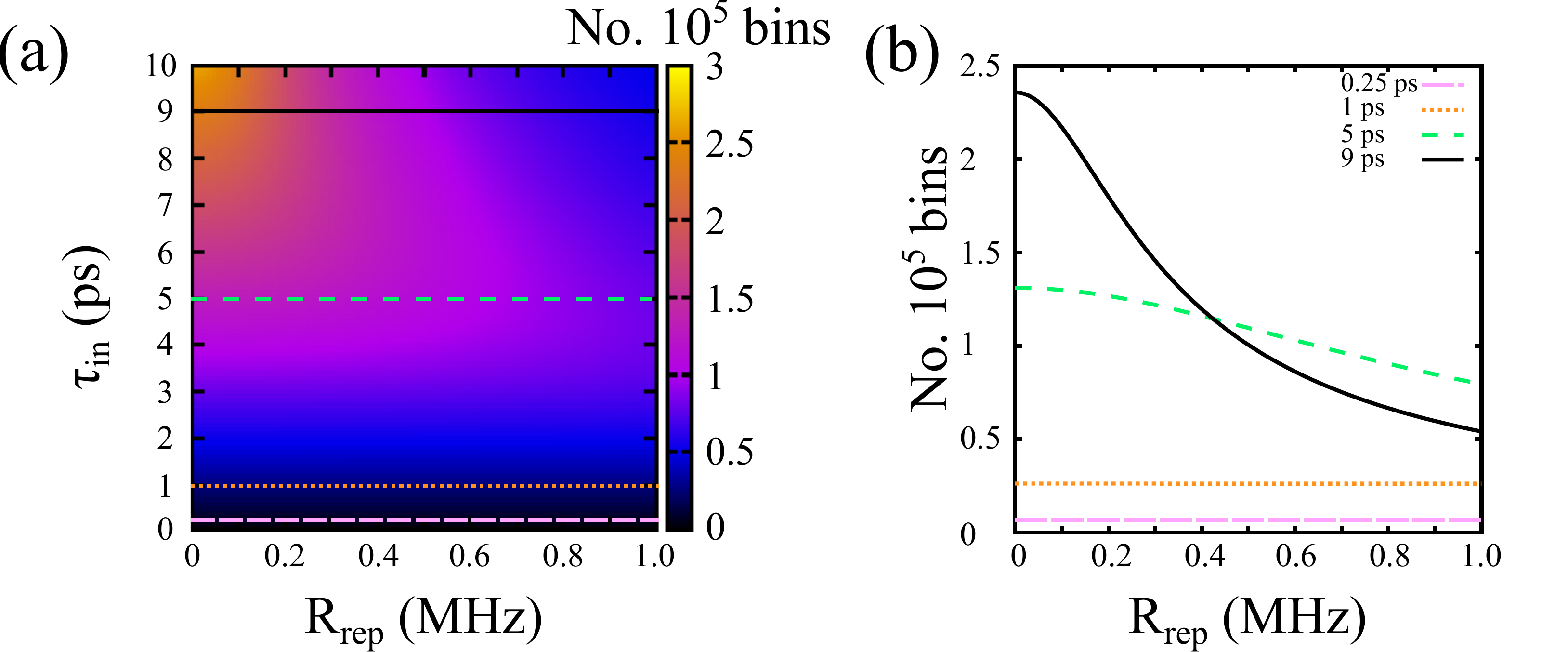}
\caption{(Color online) Maximum number of bins as limited by fiber dispersion. In (a) we consider the maximum time bin number depending on the repetition rate $R_\mathrm{Rep}$ and the input pulse duration. We find that even for low repetition rates and long pulse duration the number of time bins is bound by $\approx\, 2.5\times 10^5$ bins. In (b) we cut (a) at fixed input pulse lengths (long-dashed pink line: 250$\,$fs, dotted orange line: 1$\,$ps, short-dashed green line: 5$\,$ps, and solid black line: 9$\,$ps). For short pulse durations the decrease in the repetition rate is compensated by the increased fiber dispersion, keeping the overall number of time bins constant. For long pulses the dispersion is less pronounced, and decreasing the repetition rate helps to increase the bin number.}
\label{fig:dispersion_limit}
\end{figure}

As Eq. \eqref{eq:dispersion} applies to the FWHM of the output pulses, we have artificially halved the number of available time bins from Eq. \eqref{eq:bins} to guarantee that the different time bins can be resolved in the experiment. The results are given in Fig. \ref{fig:dispersion_limit}. In Fig. \ref{fig:dispersion_limit}(a), we plot the maximum number of available time bins depending on the repetition rate of the experiment, as well as on the input pulse duration which undergoes fiber dispersion. We consider only repetition rates below 1$\,$MHz, as they give sufficient time between two experiments, while still allowing for sufficient data rates. Even for low repetition rates (long time between two consecutive experiments) and long input pulses (low dispersion effect) we find that the maximum number of time bins is bounded by approximately $2.5\times10^5$ bins. Figure \ref{fig:dispersion_limit}(b) examines the effect of dispersion in more detail. We cut through Fig. \ref{fig:dispersion_limit}(a) at specified input pulse lengths of 250$\,$fs with the long-dashed pink line, 1$\,$ps with the dotted orange line, 5$\,$ps with the short-dashed green line and 9$\,$ps with the solid black line. It becomes clear that for short pulse durations the longer time scale between experiments at low repetition rates is fully compensated by the increased fiber dispersion and the number of available time bins remains approximately constant.

For longer input pulses, the dispersion plays a smaller role, which is why we cut the plot at $\tau_\mathrm{in, max}=10\,$ps. The number of available time bins is accordingly higher and increases even for smaller repetition rates, as the fiber dispersion is not strong enough to eliminate the advantage of longer times between experiments.

From this result, we can conclude the hard limit of available time bins in fiber-integrated TMD systems due to dispersion. For short input pulses, the available number of bins is rather limited, while longer pulses allow for a quite high photon number resolution. However, up to this point we have not considered losses in the system which will deteriorate the detected photon-number statistics (click statistics) of the input state.

\section{Limitation via Loss}
In the previous section, we considered an ideal fiber integrated system without loss and perfect detectors. Now, we consider a system including losses and finite setup transmission and detection efficiency. We investigate the ability to discriminate between different Fock states from their measured photon number statistics and also comment on the reconstruction limitations of Fock states by scanning the overlap of the adjacent Fock states and using the width of this curve as a measure for the reconstruction error.

\subsection{Model and parameters}
The action of the TMD on photon-number statistics is governed by two mechanisms: the convolution matrix that describes the distribution of $n$ photons on a finite number of $N$ bins \cite{achilles_photon-number-resolving_2004} and losses that directly deteriorate the photon-number statistics.

For this work, we use the photon-counting formula proposed by Sperling \textit{et al.} \cite{sperling_true_2012}. As their model accounts for only perfect 50:50 beam splitters in the TMD (i.e., perfectly even photon distribution among the bins), this formula is a good approximation of the expected photon-number statistics in an experiment, even for slightly different beam-splitter ratios.
With the photon-counting formula, we can formulate the convolution matrix $C$,
\begin{equation}
C_{n',k}=\left\{\begin{array}{cl} \frac{1}{N^{n'}} {N \choose k} \sum_{j=0}^{k} (-1)^j {k\choose j} (k-j)^{n'} 
, & \mbox{if }{n'}\geq k\\ 0, & \mbox{otherwise} \end{array}\right. 
\end{equation}
which gives the probability of measuring $k$ clicks when $n'$ photons impinge on $N$ detectors (in our case $N=2^b$ time bins). For more details see \cite{sperling_true_2012}.

The second mechanism is the impact of loss on the photon-number statistics. The expression for the loss matrix is well known and is given by e.g. \cite{achilles_photon-number-resolving_2004}
\begin{equation}
L_{n,n'}= {n \choose n'} \eta^{n'}(1-\eta)^{n-n'}\, .
\end{equation}
It describes the probability of retaining $n'$ photons out of $n$ with a finite efficiency $\eta$. In our case, the losses are defined by the number of beam splitters, the longest fiber length that the photons have to pass, and nonunit detection efficiencies.

Finally, the resulting click distribution after passing the TMD is given by \cite{achilles_photon-number-resolving_2004}
\begin{equation}
p_k^{(\mathrm{out})}=C_{n',k}L_{n,n'}\rho_n^{(\mathrm{in})}\, .
\end{equation}
To simplify our investigation, we consider loss and convolution as separate processes. We assume that the convolution is governed by perfect 50:50 beam splitters in the TMD, while loss affects the click statistics separately. This assumption will provide only an approximation of the realistic click statistics as losses inside the fiber network or nonperfect detection efficiencies modify the splitting ratio of the beam-splitter network \cite{achilles_photon-number-resolving_2004}. 

For experimental parameters, we consider freely available, state-of-the-art fiber-integrated components. We assume that our detectors offer a dead time of 10$\,$ns, which might be achievable for superconducting nanowire detectors in the future \cite{annunziata_reset_2010}. Furthermore, we assume fiber losses of 0.2$\,\frac{\mathrm{dB}}{\mathrm{km}}$ \cite{corning_optical_fibre_optical_2007} and a minimal loss of 0.05$\,$dB per beam splitter \cite{smolenaars_low_2016}. We will neglect the losses of the fiber splices, as they are very low compared to the splitter and fiber loss. If not otherwise specified, we assume perfect setup transmission (prior to the TMD) and detection efficiency $\eta_\mathrm{ex}$. While we assume perfect splitting ratios for the used beam splitters, strong imperfections may be accounted for. Bohmann \textit{et al.} \cite{bohmann_direct_2016} have proposed an averaging method which allows us to absorb imperfect splittings in the overall setup transmission and detection efficiency $\eta_\mathrm{ex}$. As such, our results for finite detection efficiency are also suitable to cover the case of imbalanced splitting ratios.

\subsection{Photon-number discrimination}
In the following, we consider pure Fock states as test cases for our TMD investigation. They have the advantage that they are orthonormal in the input and that the effect of modified click statistics is most pronounced.
\begin{figure}
\includegraphics[width=1.\columnwidth]{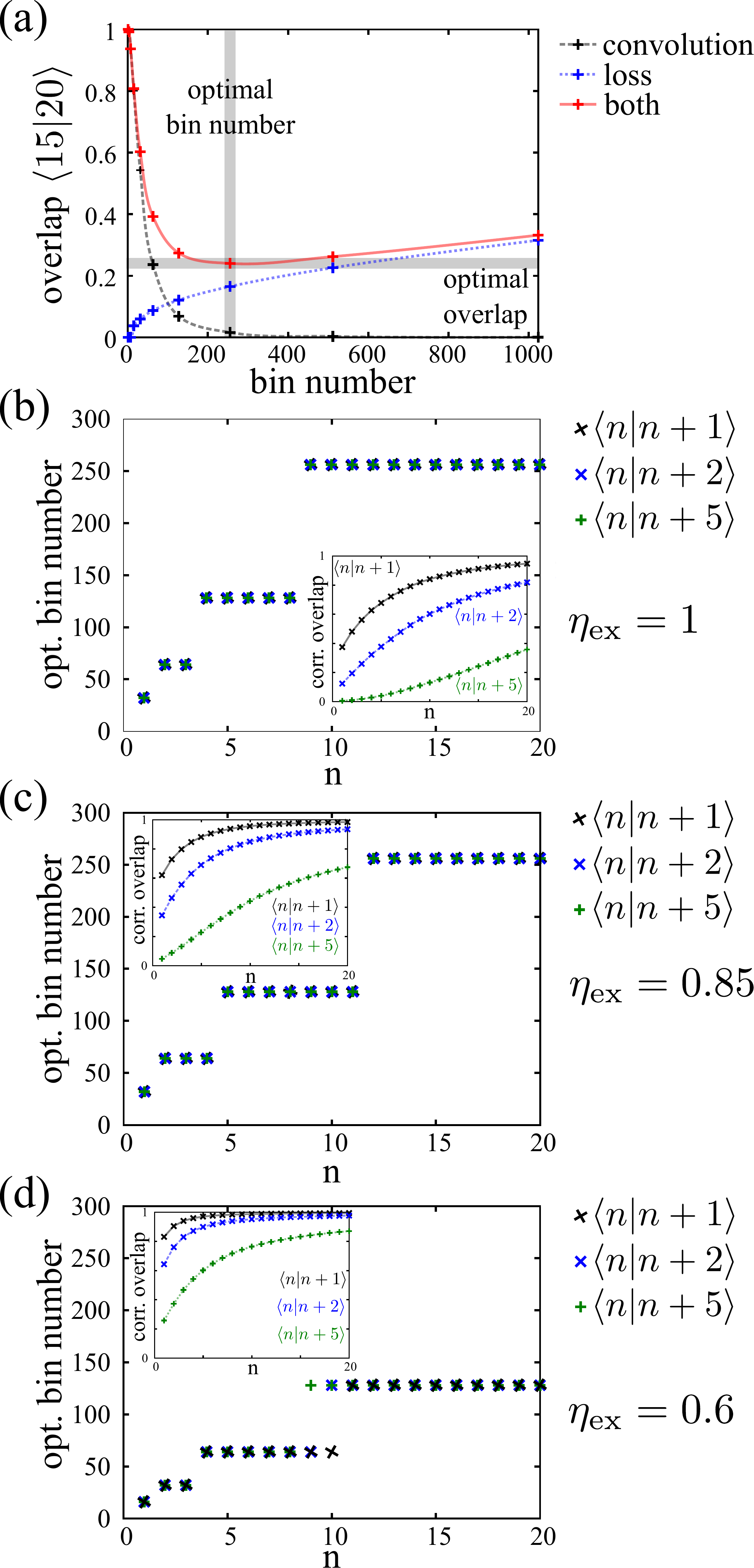}
\caption{(Color online) Overlap of the click statistics between different Fock states after passing a TMD. In (a) we consider the different contributions to the final overlap separately for the example $\bra{15}20\rangle$. For small bin numbers, the overlap is governed by the convolution matrix (dashed line, black pluses); for high bin numbers it is governed by the losses (dotted line, blue pluses). The minimum of the curve that considers both effects (solid line, red pluses) gives the optimal parameters for minimal overlap, marked in gray. In (b)-(d), we extract the optimal bin number and overlap (curves in the insets) for different detection and setup efficiencies $\eta$. The additional losses deteriorate the click statistics quite severely, such that for realistic setup and detection efficiencies ($\eta=0.8$) it holds no advantage to implement TMDs larger than 256 bins. The faint lines hold no physical meaning and are provided as a guide to the eye.}
\label{fig:opt_overlap}
\end{figure}

To quantify the effect of TMD measurements on the input statistics, we first consider the effects of the convolution and loss separately in Fig. \ref{fig:opt_overlap}(a).
In black (dashed line), we plot the overlap of the click statistics of the two Fock states $\bra{15}20\rangle$ as an example. As expected \cite{sperling_true_2012}, the overlap decreases to higher bin numbers as the effect of the convolution decreases and the click statistics approximate the true photon-number statistics. The opposite is true for the losses, as shown in blue (dotted line). As the number of beam splitters and fiber length increase, so do the losses for more bins. Accordingly, the click statistics are washed out, and the overlap increases.
Therefore, when considering both convolution and loss effects (plotted in red, solid line), we find an optimal bin number where the overlap between the two Fock states after passing the TMD is minimal.

In Figs. \ref{fig:opt_overlap}(b)-\ref{fig:opt_overlap}(d), we extract the optimal TMD parameters for different input Fock states and detection efficiencies. We consider neighboring (black tilted crosses) Fock states, next-nearest neighbours (blue crosses), and states with with four numbers in between (green pluses). We find that even for the ideal case of perfect setup transmission and detection efficiency in Fig. \ref{fig:opt_overlap}(b), the optimal bin number for all three cases at high photon numbers does not exceed 256 bins. The associated overlap for the optimal bin numbers (plotted as insets) is a monotonically increasing function of the photon number and reaches 1 asymptotically. Realistically, this curve will have to be truncated at a critical overlap value that depends on the application, the robustness of the reconstruction algorithm of choice, and also the tolerable measurement time and statistical errors in the experiment.

Adding loss that accounts for imperfect setup transmission and detection efficiency in Figs. \ref{fig:opt_overlap}(c) and \ref{fig:opt_overlap}(d) deteriorates the click statistics quite drastically. In this case, the optimum bin numbers are lower than for the ideal case, while the overlap curves approach 1 faster. In the case of a realistic, overall experiment efficiency of $\eta_\mathrm{ex}=0.85$, it becomes clear that for our figure of merit the optimal TMD size does not exceed $256=2^8$ bins. 


\subsection{Photon-number reconstruction}
In the last section, we saw there is no advantage to build TMDs bigger than 256 time bins when using the setup to discriminate between Fock states. However, this might not be the only aim of a quasiphoton-number-resolved detection. Another significant task is to reconstruct the impinging states on the TMD. To this aim a lot of reconstruction algorithms have been proposed and implemented (e.g., see \cite{zambra_experimental_2005, dodonov_engineering_2006, allevi_photon-number_2010}). In this paper, we do not want to comment on the advantages or disadvantages of the particular reconstruction methods and only want to  infer an error bar from the measured statistics that will affect the precision of the reconstruction.
\begin{figure}
\includegraphics[width=1.\columnwidth]{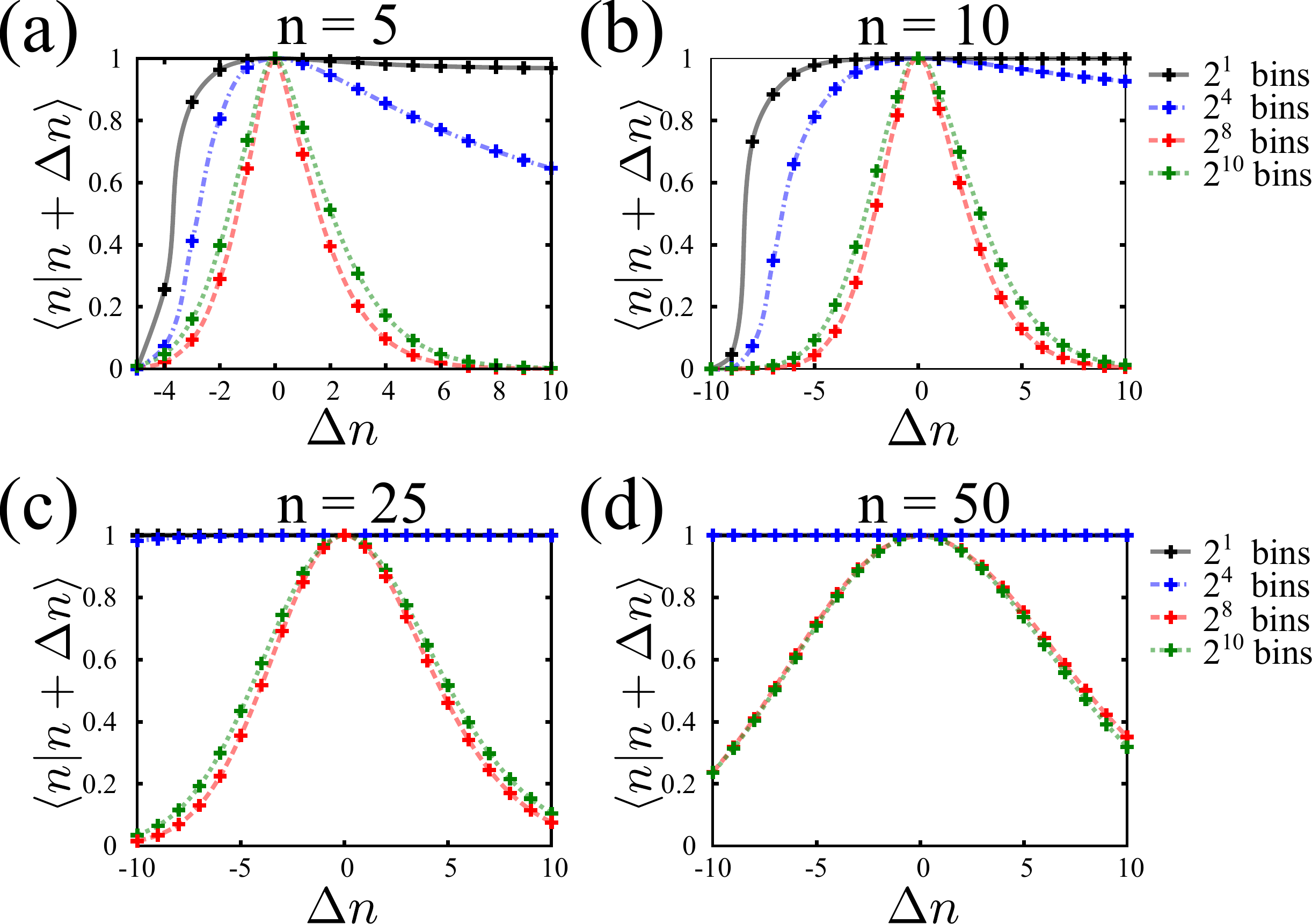}
\caption{(Color online) For different, but fixed Fock state inputs $\ket{n}$, we calculate the overlap to the adjacent $\pm$10 Fock states in the click statistics after passing TMDs of different size. The width of the overlap curve gives a measure with which precision a Fock state can be measured and is therefore a criterion for the resolution of the measurement method. Even for very small input Fock states [$n=5$ in (a)] large TMDs ($2^{10}$ bins in green, dotted line) do not perfectly resolve the input state. For larger states [e.g., $n=50$ in (d)], the resolution goes down drastically. From this we conclude that TMDs are not sufficient to verify large quantum states directly. The faint lines in the plots hold no physical meaning and are provided as a guide to the eye.}
\label{fig:reconstruction}
\end{figure}

To this aim, we regard a pure Fock state $\ket{n}$ impinging on our detector without excess loss ($\eta_\mathrm{ex}=1$) and calculate the overlap of the click statistics after the TMD for the 20 adjacent Fock states $\bra{n-10}n\rangle$ to $\bra{n}n+10\rangle$. The results for different input states $\ket{n}$ and different bin numbers are depicted in Fig. \ref{fig:reconstruction}.

As an example, let us consider figure \ref{fig:reconstruction}(a). We encoded the overlap of the click statistics after passing TMDs of different sizes both in color and line style. Consider the red dashed curve for passing a 256-bin TMD. The curve peaks at $\Delta n=0$, as we send the same state into the TMD. It is slightly asymmetric with respect to $\Delta n=0$, as the binomial coefficients that govern the overlap of the click statistics are different for the higher-and-lower-photon-number cases. This effect evens out to higher input Fock states [see Fig. \ref{fig:reconstruction}(d)] as the relative difference in the input photon number decreases. 

As we increase the photon number of the input Fock states, the overlap curve between the click statistics becomes broader for all considered TMD sizes. This is expected due to the increased impact of the convolution matrix for higher photon numbers. However, it also becomes clear that the improvement of larger TMDs for large input states is not very pronounced. Especially for Fig. \ref{fig:reconstruction}(d), the improvement in the width of the overlap curve between 256 and 1024 bins does not really justify the experimental effort it takes to fabricate a high quality TMD of that size. Furthermore, one has to consider the impact of finite detection efficiencies and setup transmission. We saw in the previous section that the increased losses only deteriorate the statistics further and cause building large TMDs without to have no real advantage.

\section{Conclusion}
In conclusion we investigated the limitations of photon-number-resolved measurements by time-multiplexed detection. We discussed the fundamental limit of the device as given by its dispersive fiber-integrated design. Furthermore, we considered the combined effect of losses and convolution in the context of photon-number-discrimination tasks, as well as for photon-number-state reconstruction. Both cases show that building large TMDs is not advantageous since losses deteriorate the photon-number statistics faster than the effect of the convolution matrix diminishes. As a recommendation based on realistic experimental 
figures of merit, we suggest using 256 bin devices, as they provide both moderate losses and moderate photon-number resolution.
\vspace{0.5cm}

\textbf{Acknowledgments}\\
The authors would like to thank M. Bohmann for valuable discussions concerning the photon counting formula. This work was supported by European Union Grant No. 665148 (QCUMbER). T.J.B. acknowledges support from the DFG (Deutsche Forschungsgesellschaft) under Grant SFB/TRR 142..

\bibliography{Limits-on-TMD}

\begin{thebibliography}{31}%
\makeatletter
\providecommand \@ifxundefined [1]{%
 \@ifx{#1\undefined}
}%
\providecommand \@ifnum [1]{%
 \ifnum #1\expandafter \@firstoftwo
 \else \expandafter \@secondoftwo
 \fi
}%
\providecommand \@ifx [1]{%
 \ifx #1\expandafter \@firstoftwo
 \else \expandafter \@secondoftwo
 \fi
}%
\providecommand \natexlab [1]{#1}%
\providecommand \enquote  [1]{``#1''}%
\providecommand \bibnamefont  [1]{#1}%
\providecommand \bibfnamefont [1]{#1}%
\providecommand \citenamefont [1]{#1}%
\providecommand \href@noop [0]{\@secondoftwo}%
\providecommand \href [0]{\begingroup \@sanitize@url \@href}%
\providecommand \@href[1]{\@@startlink{#1}\@@href}%
\providecommand \@@href[1]{\endgroup#1\@@endlink}%
\providecommand \@sanitize@url [0]{\catcode `\\12\catcode `\$12\catcode
  `\&12\catcode `\#12\catcode `\^12\catcode `\_12\catcode `\%12\relax}%
\providecommand \@@startlink[1]{}%
\providecommand \@@endlink[0]{}%
\providecommand \url  [0]{\begingroup\@sanitize@url \@url }%
\providecommand \@url [1]{\endgroup\@href {#1}{\urlprefix }}%
\providecommand \urlprefix  [0]{URL }%
\providecommand \Eprint [0]{\href }%
\providecommand \doibase [0]{http://dx.doi.org/}%
\providecommand \selectlanguage [0]{\@gobble}%
\providecommand \bibinfo  [0]{\@secondoftwo}%
\providecommand \bibfield  [0]{\@secondoftwo}%
\providecommand \translation [1]{[#1]}%
\providecommand \BibitemOpen [0]{}%
\providecommand \bibitemStop [0]{}%
\providecommand \bibitemNoStop [0]{.\EOS\space}%
\providecommand \EOS [0]{\spacefactor3000\relax}%
\providecommand \BibitemShut  [1]{\csname bibitem#1\endcsname}%
\let\auto@bib@innerbib\@empty
\bibitem [{\citenamefont {Mosley}\ \emph {et~al.}(2008)\citenamefont {Mosley},
  \citenamefont {Lundeen}, \citenamefont {Smith}, \citenamefont {Wasylczyk},
  \citenamefont {U’Ren}, \citenamefont {Silberhorn},\ and\ \citenamefont
  {Walmsley}}]{mosley_heralded_2008}%
  \BibitemOpen
  \bibfield  {author} {\bibinfo {author} {\bibfnamefont {P.~J.}\ \bibnamefont
  {Mosley}}, \bibinfo {author} {\bibfnamefont {J.~S.}\ \bibnamefont {Lundeen}},
  \bibinfo {author} {\bibfnamefont {B.~J.}\ \bibnamefont {Smith}}, \bibinfo
  {author} {\bibfnamefont {P.}~\bibnamefont {Wasylczyk}}, \bibinfo {author}
  {\bibfnamefont {A.~B.}\ \bibnamefont {U’Ren}}, \bibinfo {author}
  {\bibfnamefont {C.}~\bibnamefont {Silberhorn}}, \ and\ \bibinfo {author}
  {\bibfnamefont {I.~A.}\ \bibnamefont {Walmsley}},\ }\href {\doibase
  10.1103/PhysRevLett.100.133601} {\bibfield  {journal} {\bibinfo  {journal}
  {Physical Review Letters}\ }\textbf {\bibinfo {volume} {100}},\ \bibinfo
  {pages} {133601} (\bibinfo {year} {2008})}\BibitemShut {NoStop}%
\bibitem [{\citenamefont {Eckstein}\ \emph {et~al.}(2011)\citenamefont
  {Eckstein}, \citenamefont {Christ}, \citenamefont {Mosley},\ and\
  \citenamefont {Silberhorn}}]{eckstein_highly_2011}%
  \BibitemOpen
  \bibfield  {author} {\bibinfo {author} {\bibfnamefont {A.}~\bibnamefont
  {Eckstein}}, \bibinfo {author} {\bibfnamefont {A.}~\bibnamefont {Christ}},
  \bibinfo {author} {\bibfnamefont {P.~J.}\ \bibnamefont {Mosley}}, \ and\
  \bibinfo {author} {\bibfnamefont {C.}~\bibnamefont {Silberhorn}},\ }\href
  {\doibase 10.1103/PhysRevLett.106.013603} {\bibfield  {journal} {\bibinfo
  {journal} {Physical Review Letters}\ }\textbf {\bibinfo {volume} {106}},\
  \bibinfo {pages} {013603} (\bibinfo {year} {2011})}\BibitemShut {NoStop}%
\bibitem [{\citenamefont {Harder}\ \emph {et~al.}(2013)\citenamefont {Harder},
  \citenamefont {Ansari}, \citenamefont {Brecht}, \citenamefont {Dirmeier},
  \citenamefont {Marquardt},\ and\ \citenamefont
  {Silberhorn}}]{harder_optimized_2013}%
  \BibitemOpen
  \bibfield  {author} {\bibinfo {author} {\bibfnamefont {G.}~\bibnamefont
  {Harder}}, \bibinfo {author} {\bibfnamefont {V.}~\bibnamefont {Ansari}},
  \bibinfo {author} {\bibfnamefont {B.}~\bibnamefont {Brecht}}, \bibinfo
  {author} {\bibfnamefont {T.}~\bibnamefont {Dirmeier}}, \bibinfo {author}
  {\bibfnamefont {C.}~\bibnamefont {Marquardt}}, \ and\ \bibinfo {author}
  {\bibfnamefont {C.}~\bibnamefont {Silberhorn}},\ }\href {\doibase
  10.1364/OE.21.013975} {\bibfield  {journal} {\bibinfo  {journal} {Optics
  Express}\ }\textbf {\bibinfo {volume} {21}},\ \bibinfo {pages} {13975}
  (\bibinfo {year} {2013})}\BibitemShut {NoStop}%
\bibitem [{\citenamefont {Collins}\ \emph {et~al.}(2013)\citenamefont
  {Collins}, \citenamefont {Xiong}, \citenamefont {Rey}, \citenamefont {Vo},
  \citenamefont {He}, \citenamefont {Shahnia}, \citenamefont {Reardon},
  \citenamefont {Krauss}, \citenamefont {Steel}, \citenamefont {Clark},\ and\
  \citenamefont {Eggleton}}]{collins_integrated_2013}%
  \BibitemOpen
  \bibfield  {author} {\bibinfo {author} {\bibfnamefont {M.~J.}\ \bibnamefont
  {Collins}}, \bibinfo {author} {\bibfnamefont {C.}~\bibnamefont {Xiong}},
  \bibinfo {author} {\bibfnamefont {I.~H.}\ \bibnamefont {Rey}}, \bibinfo
  {author} {\bibfnamefont {T.~D.}\ \bibnamefont {Vo}}, \bibinfo {author}
  {\bibfnamefont {J.}~\bibnamefont {He}}, \bibinfo {author} {\bibfnamefont
  {S.}~\bibnamefont {Shahnia}}, \bibinfo {author} {\bibfnamefont
  {C.}~\bibnamefont {Reardon}}, \bibinfo {author} {\bibfnamefont {T.~F.}\
  \bibnamefont {Krauss}}, \bibinfo {author} {\bibfnamefont {M.~J.}\
  \bibnamefont {Steel}}, \bibinfo {author} {\bibfnamefont {A.~S.}\ \bibnamefont
  {Clark}}, \ and\ \bibinfo {author} {\bibfnamefont {B.~J.}\ \bibnamefont
  {Eggleton}},\ }\href {\doibase 10.1038/ncomms3582} {\bibfield  {journal}
  {\bibinfo  {journal} {Nature Communications}\ }\textbf {\bibinfo {volume}
  {4}},\ \bibinfo {pages} {2582} (\bibinfo {year} {2013})}\BibitemShut
  {NoStop}%
\bibitem [{\citenamefont {Kumar}\ \emph {et~al.}(2014)\citenamefont {Kumar},
  \citenamefont {Ong}, \citenamefont {Savanier},\ and\ \citenamefont
  {Mookherjea}}]{kumar_controlling_2014}%
  \BibitemOpen
  \bibfield  {author} {\bibinfo {author} {\bibfnamefont {R.}~\bibnamefont
  {Kumar}}, \bibinfo {author} {\bibfnamefont {J.~R.}\ \bibnamefont {Ong}},
  \bibinfo {author} {\bibfnamefont {M.}~\bibnamefont {Savanier}}, \ and\
  \bibinfo {author} {\bibfnamefont {S.}~\bibnamefont {Mookherjea}},\ }\href
  {\doibase 10.1038/ncomms6489} {\bibfield  {journal} {\bibinfo  {journal}
  {Nature Communications}\ }\textbf {\bibinfo {volume} {5}},\ \bibinfo {pages}
  {5489} (\bibinfo {year} {2014})}\BibitemShut {NoStop}%
\bibitem [{\citenamefont {Bruno}\ \emph {et~al.}(2014)\citenamefont {Bruno},
  \citenamefont {Martin}, \citenamefont {Guerreiro}, \citenamefont
  {Sanguinetti},\ and\ \citenamefont {Thew}}]{bruno_pulsed_2014}%
  \BibitemOpen
  \bibfield  {author} {\bibinfo {author} {\bibfnamefont {N.}~\bibnamefont
  {Bruno}}, \bibinfo {author} {\bibfnamefont {A.}~\bibnamefont {Martin}},
  \bibinfo {author} {\bibfnamefont {T.}~\bibnamefont {Guerreiro}}, \bibinfo
  {author} {\bibfnamefont {B.}~\bibnamefont {Sanguinetti}}, \ and\ \bibinfo
  {author} {\bibfnamefont {R.~T.}\ \bibnamefont {Thew}},\ }\href {\doibase
  10.1364/OE.22.017246} {\bibfield  {journal} {\bibinfo  {journal} {Optics
  Express}\ }\textbf {\bibinfo {volume} {22}},\ \bibinfo {pages} {17246}
  (\bibinfo {year} {2014})}\BibitemShut {NoStop}%
\bibitem [{\citenamefont {Jin}\ \emph {et~al.}(2015)\citenamefont {Jin},
  \citenamefont {Fujiwara}, \citenamefont {Yamashita}, \citenamefont {Miki},
  \citenamefont {Terai}, \citenamefont {Wang}, \citenamefont {Wakui},
  \citenamefont {Shimizu},\ and\ \citenamefont {Sasaki}}]{jin_efficient_2015}%
  \BibitemOpen
  \bibfield  {author} {\bibinfo {author} {\bibfnamefont {R.-B.}\ \bibnamefont
  {Jin}}, \bibinfo {author} {\bibfnamefont {M.}~\bibnamefont {Fujiwara}},
  \bibinfo {author} {\bibfnamefont {T.}~\bibnamefont {Yamashita}}, \bibinfo
  {author} {\bibfnamefont {S.}~\bibnamefont {Miki}}, \bibinfo {author}
  {\bibfnamefont {H.}~\bibnamefont {Terai}}, \bibinfo {author} {\bibfnamefont
  {Z.}~\bibnamefont {Wang}}, \bibinfo {author} {\bibfnamefont {K.}~\bibnamefont
  {Wakui}}, \bibinfo {author} {\bibfnamefont {R.}~\bibnamefont {Shimizu}}, \
  and\ \bibinfo {author} {\bibfnamefont {M.}~\bibnamefont {Sasaki}},\ }\href
  {\doibase 10.1016/j.optcom.2014.09.051} {\bibfield  {journal} {\bibinfo
  {journal} {Optics Communications}\ }\textbf {\bibinfo {volume} {336}},\
  \bibinfo {pages} {47} (\bibinfo {year} {2015})}\BibitemShut {NoStop}%
\bibitem [{\citenamefont {Förtsch}\ \emph {et~al.}(2015)\citenamefont
  {Förtsch}, \citenamefont {Schunk}, \citenamefont {Fürst}, \citenamefont
  {Strekalov}, \citenamefont {Gerrits}, \citenamefont {Stevens}, \citenamefont
  {Sedlmeir}, \citenamefont {Schwefel}, \citenamefont {Nam}, \citenamefont
  {Leuchs},\ and\ \citenamefont {Marquardt}}]{fortsch_highly_2015}%
  \BibitemOpen
  \bibfield  {author} {\bibinfo {author} {\bibfnamefont {M.}~\bibnamefont
  {Förtsch}}, \bibinfo {author} {\bibfnamefont {G.}~\bibnamefont {Schunk}},
  \bibinfo {author} {\bibfnamefont {J.~U.}\ \bibnamefont {Fürst}}, \bibinfo
  {author} {\bibfnamefont {D.}~\bibnamefont {Strekalov}}, \bibinfo {author}
  {\bibfnamefont {T.}~\bibnamefont {Gerrits}}, \bibinfo {author} {\bibfnamefont
  {M.~J.}\ \bibnamefont {Stevens}}, \bibinfo {author} {\bibfnamefont
  {F.}~\bibnamefont {Sedlmeir}}, \bibinfo {author} {\bibfnamefont {H.~G.~L.}\
  \bibnamefont {Schwefel}}, \bibinfo {author} {\bibfnamefont {S.~W.}\
  \bibnamefont {Nam}}, \bibinfo {author} {\bibfnamefont {G.}~\bibnamefont
  {Leuchs}}, \ and\ \bibinfo {author} {\bibfnamefont {C.}~\bibnamefont
  {Marquardt}},\ }\href {\doibase 10.1103/PhysRevA.91.023812} {\bibfield
  {journal} {\bibinfo  {journal} {Physical Review A}\ }\textbf {\bibinfo
  {volume} {91}},\ \bibinfo {pages} {023812} (\bibinfo {year}
  {2015})}\BibitemShut {NoStop}%
\bibitem [{\citenamefont {Zielnicki}\ \emph {et~al.}(2015)\citenamefont
  {Zielnicki}, \citenamefont {Garay-Palmett}, \citenamefont {Dirks},
  \citenamefont {U’Ren},\ and\ \citenamefont
  {Kwiat}}]{zielnicki_engineering_2015}%
  \BibitemOpen
  \bibfield  {author} {\bibinfo {author} {\bibfnamefont {K.}~\bibnamefont
  {Zielnicki}}, \bibinfo {author} {\bibfnamefont {K.}~\bibnamefont
  {Garay-Palmett}}, \bibinfo {author} {\bibfnamefont {R.}~\bibnamefont
  {Dirks}}, \bibinfo {author} {\bibfnamefont {A.~B.}\ \bibnamefont {U’Ren}},
  \ and\ \bibinfo {author} {\bibfnamefont {P.~G.}\ \bibnamefont {Kwiat}},\
  }\href {\doibase 10.1364/OE.23.007894} {\bibfield  {journal} {\bibinfo
  {journal} {Optics Express}\ }\textbf {\bibinfo {volume} {23}},\ \bibinfo
  {pages} {7894} (\bibinfo {year} {2015})}\BibitemShut {NoStop}%
\bibitem [{\citenamefont {Harder}\ \emph {et~al.}(2016)\citenamefont {Harder},
  \citenamefont {Bartley}, \citenamefont {Lita}, \citenamefont {Nam},
  \citenamefont {Gerrits},\ and\ \citenamefont
  {Silberhorn}}]{harder_single-mode_2016}%
  \BibitemOpen
  \bibfield  {author} {\bibinfo {author} {\bibfnamefont {G.}~\bibnamefont
  {Harder}}, \bibinfo {author} {\bibfnamefont {T.~J.}\ \bibnamefont {Bartley}},
  \bibinfo {author} {\bibfnamefont {A.~E.}\ \bibnamefont {Lita}}, \bibinfo
  {author} {\bibfnamefont {S.~W.}\ \bibnamefont {Nam}}, \bibinfo {author}
  {\bibfnamefont {T.}~\bibnamefont {Gerrits}}, \ and\ \bibinfo {author}
  {\bibfnamefont {C.}~\bibnamefont {Silberhorn}},\ }\href {\doibase
  10.1103/PhysRevLett.116.143601} {\bibfield  {journal} {\bibinfo  {journal}
  {Physical Review Letters}\ }\textbf {\bibinfo {volume} {116}},\ \bibinfo
  {pages} {143601} (\bibinfo {year} {2016})}\BibitemShut {NoStop}%
\bibitem [{\citenamefont {Lita}\ \emph {et~al.}(2008)\citenamefont {Lita},
  \citenamefont {Miller},\ and\ \citenamefont {Nam}}]{lita_counting_2008}%
  \BibitemOpen
  \bibfield  {author} {\bibinfo {author} {\bibfnamefont {A.~E.}\ \bibnamefont
  {Lita}}, \bibinfo {author} {\bibfnamefont {A.~J.}\ \bibnamefont {Miller}}, \
  and\ \bibinfo {author} {\bibfnamefont {S.~W.}\ \bibnamefont {Nam}},\ }\href
  {\doibase 10.1364/OE.16.003032} {\bibfield  {journal} {\bibinfo  {journal}
  {Optics Express}\ }\textbf {\bibinfo {volume} {16}},\ \bibinfo {pages} {3032}
  (\bibinfo {year} {2008})}\BibitemShut {NoStop}%
\bibitem [{\citenamefont {Marsili}\ \emph {et~al.}(2013)\citenamefont
  {Marsili}, \citenamefont {Verma}, \citenamefont {Stern}, \citenamefont
  {Harrington}, \citenamefont {Lita}, \citenamefont {Gerrits}, \citenamefont
  {Vayshenker}, \citenamefont {Baek}, \citenamefont {Shaw}, \citenamefont
  {Mirin},\ and\ \citenamefont {Nam}}]{marsili_detecting_2013}%
  \BibitemOpen
  \bibfield  {author} {\bibinfo {author} {\bibfnamefont {F.}~\bibnamefont
  {Marsili}}, \bibinfo {author} {\bibfnamefont {V.~B.}\ \bibnamefont {Verma}},
  \bibinfo {author} {\bibfnamefont {J.~A.}\ \bibnamefont {Stern}}, \bibinfo
  {author} {\bibfnamefont {S.}~\bibnamefont {Harrington}}, \bibinfo {author}
  {\bibfnamefont {A.~E.}\ \bibnamefont {Lita}}, \bibinfo {author}
  {\bibfnamefont {T.}~\bibnamefont {Gerrits}}, \bibinfo {author} {\bibfnamefont
  {I.}~\bibnamefont {Vayshenker}}, \bibinfo {author} {\bibfnamefont
  {B.}~\bibnamefont {Baek}}, \bibinfo {author} {\bibfnamefont {M.~D.}\
  \bibnamefont {Shaw}}, \bibinfo {author} {\bibfnamefont {R.~P.}\ \bibnamefont
  {Mirin}}, \ and\ \bibinfo {author} {\bibfnamefont {S.~W.}\ \bibnamefont
  {Nam}},\ }\href {\doibase 10.1038/nphoton.2013.13} {\bibfield  {journal}
  {\bibinfo  {journal} {Nature Photonics}\ }\textbf {\bibinfo {volume} {7}},\
  \bibinfo {pages} {210} (\bibinfo {year} {2013})}\BibitemShut {NoStop}%
\bibitem [{\citenamefont {Fukuda}\ \emph {et~al.}(2011)\citenamefont {Fukuda},
  \citenamefont {Fujii}, \citenamefont {Numata}, \citenamefont {Amemiya},
  \citenamefont {Yoshizawa}, \citenamefont {Tsuchida}, \citenamefont {Fujino},
  \citenamefont {Ishii}, \citenamefont {Itatani}, \citenamefont {Inoue},\ and\
  \citenamefont {Zama}}]{fukuda_titanium-based_2011}%
  \BibitemOpen
  \bibfield  {author} {\bibinfo {author} {\bibfnamefont {D.}~\bibnamefont
  {Fukuda}}, \bibinfo {author} {\bibfnamefont {G.}~\bibnamefont {Fujii}},
  \bibinfo {author} {\bibfnamefont {T.}~\bibnamefont {Numata}}, \bibinfo
  {author} {\bibfnamefont {K.}~\bibnamefont {Amemiya}}, \bibinfo {author}
  {\bibfnamefont {A.}~\bibnamefont {Yoshizawa}}, \bibinfo {author}
  {\bibfnamefont {H.}~\bibnamefont {Tsuchida}}, \bibinfo {author}
  {\bibfnamefont {H.}~\bibnamefont {Fujino}}, \bibinfo {author} {\bibfnamefont
  {H.}~\bibnamefont {Ishii}}, \bibinfo {author} {\bibfnamefont
  {T.}~\bibnamefont {Itatani}}, \bibinfo {author} {\bibfnamefont
  {S.}~\bibnamefont {Inoue}}, \ and\ \bibinfo {author} {\bibfnamefont
  {T.}~\bibnamefont {Zama}},\ }\href {\doibase 10.1364/OE.19.000870} {\bibfield
   {journal} {\bibinfo  {journal} {Optics Express}\ }\textbf {\bibinfo {volume}
  {19}},\ \bibinfo {pages} {870} (\bibinfo {year} {2011})}\BibitemShut
  {NoStop}%
\bibitem [{\citenamefont {Gerrits}\ \emph {et~al.}(2012)\citenamefont
  {Gerrits}, \citenamefont {Calkins}, \citenamefont {Tomlin}, \citenamefont
  {Lita}, \citenamefont {Migdall}, \citenamefont {Mirin},\ and\ \citenamefont
  {Nam}}]{gerrits_extending_2012}%
  \BibitemOpen
  \bibfield  {author} {\bibinfo {author} {\bibfnamefont {T.}~\bibnamefont
  {Gerrits}}, \bibinfo {author} {\bibfnamefont {B.}~\bibnamefont {Calkins}},
  \bibinfo {author} {\bibfnamefont {N.}~\bibnamefont {Tomlin}}, \bibinfo
  {author} {\bibfnamefont {A.~E.}\ \bibnamefont {Lita}}, \bibinfo {author}
  {\bibfnamefont {A.}~\bibnamefont {Migdall}}, \bibinfo {author} {\bibfnamefont
  {R.}~\bibnamefont {Mirin}}, \ and\ \bibinfo {author} {\bibfnamefont {S.~W.}\
  \bibnamefont {Nam}},\ }\href {\doibase 10.1364/OE.20.023798} {\bibfield
  {journal} {\bibinfo  {journal} {Optics Express}\ }\textbf {\bibinfo {volume}
  {20}},\ \bibinfo {pages} {23798} (\bibinfo {year} {2012})}\BibitemShut
  {NoStop}%
\bibitem [{\citenamefont {Humphreys}\ \emph {et~al.}(2015)\citenamefont
  {Humphreys}, \citenamefont {Metcalf}, \citenamefont {Gerrits}, \citenamefont
  {Hiemstra}, \citenamefont {Lita}, \citenamefont {Nunn}, \citenamefont {Nam},
  \citenamefont {Datta}, \citenamefont {Kolthammer},\ and\ \citenamefont
  {Walmsley}}]{humphreys_tomography_2015}%
  \BibitemOpen
  \bibfield  {author} {\bibinfo {author} {\bibfnamefont {P.~C.}\ \bibnamefont
  {Humphreys}}, \bibinfo {author} {\bibfnamefont {B.~J.}\ \bibnamefont
  {Metcalf}}, \bibinfo {author} {\bibfnamefont {T.}~\bibnamefont {Gerrits}},
  \bibinfo {author} {\bibfnamefont {T.}~\bibnamefont {Hiemstra}}, \bibinfo
  {author} {\bibfnamefont {A.~E.}\ \bibnamefont {Lita}}, \bibinfo {author}
  {\bibfnamefont {J.}~\bibnamefont {Nunn}}, \bibinfo {author} {\bibfnamefont
  {S.~W.}\ \bibnamefont {Nam}}, \bibinfo {author} {\bibfnamefont
  {A.}~\bibnamefont {Datta}}, \bibinfo {author} {\bibfnamefont {W.~S.}\
  \bibnamefont {Kolthammer}}, \ and\ \bibinfo {author} {\bibfnamefont {I.~A.}\
  \bibnamefont {Walmsley}},\ }\href {\doibase 10.1088/1367-2630/17/10/103044}
  {\bibfield  {journal} {\bibinfo  {journal} {New Journal of Physics}\ }\textbf
  {\bibinfo {volume} {17}},\ \bibinfo {pages} {103044} (\bibinfo {year}
  {2015})}\BibitemShut {NoStop}%
\bibitem [{\citenamefont {Achilles}\ \emph {et~al.}(2003)\citenamefont
  {Achilles}, \citenamefont {Silberhorn}, \citenamefont {Śliwa}, \citenamefont
  {Banaszek},\ and\ \citenamefont {Walmsley}}]{achilles_fiber-assisted_2003}%
  \BibitemOpen
  \bibfield  {author} {\bibinfo {author} {\bibfnamefont {D.}~\bibnamefont
  {Achilles}}, \bibinfo {author} {\bibfnamefont {C.}~\bibnamefont
  {Silberhorn}}, \bibinfo {author} {\bibfnamefont {C.}~\bibnamefont {Śliwa}},
  \bibinfo {author} {\bibfnamefont {K.}~\bibnamefont {Banaszek}}, \ and\
  \bibinfo {author} {\bibfnamefont {I.~A.}\ \bibnamefont {Walmsley}},\ }\href
  {\doibase 10.1364/OL.28.002387} {\bibfield  {journal} {\bibinfo  {journal}
  {Optics Letters}\ }\textbf {\bibinfo {volume} {28}},\ \bibinfo {pages} {2387}
  (\bibinfo {year} {2003})}\BibitemShut {NoStop}%
\bibitem [{\citenamefont {Řeháček}\ \emph {et~al.}(2003)\citenamefont
  {Řeháček}, \citenamefont {Hradil}, \citenamefont {Haderka}, \citenamefont
  {Peřina},\ and\ \citenamefont {Hamar}}]{rehacek_multiple-photon_2003}%
  \BibitemOpen
  \bibfield  {author} {\bibinfo {author} {\bibfnamefont {J.}~\bibnamefont
  {Řeháček}}, \bibinfo {author} {\bibfnamefont {Z.}~\bibnamefont {Hradil}},
  \bibinfo {author} {\bibfnamefont {O.}~\bibnamefont {Haderka}}, \bibinfo
  {author} {\bibfnamefont {J.}~\bibnamefont {Peřina}}, \ and\ \bibinfo
  {author} {\bibfnamefont {M.}~\bibnamefont {Hamar}},\ }\href {\doibase
  10.1103/PhysRevA.67.061801} {\bibfield  {journal} {\bibinfo  {journal}
  {Physical Review A}\ }\textbf {\bibinfo {volume} {67}},\ \bibinfo {pages}
  {061801} (\bibinfo {year} {2003})}\BibitemShut {NoStop}%
\bibitem [{\citenamefont {Fitch}\ \emph {et~al.}(2003)\citenamefont {Fitch},
  \citenamefont {Jacobs}, \citenamefont {Pittman},\ and\ \citenamefont
  {Franson}}]{fitch_photon-number_2003}%
  \BibitemOpen
  \bibfield  {author} {\bibinfo {author} {\bibfnamefont {M.~J.}\ \bibnamefont
  {Fitch}}, \bibinfo {author} {\bibfnamefont {B.~C.}\ \bibnamefont {Jacobs}},
  \bibinfo {author} {\bibfnamefont {T.~B.}\ \bibnamefont {Pittman}}, \ and\
  \bibinfo {author} {\bibfnamefont {J.~D.}\ \bibnamefont {Franson}},\ }\href
  {\doibase 10.1103/PhysRevA.68.043814} {\bibfield  {journal} {\bibinfo
  {journal} {Physical Review A}\ }\textbf {\bibinfo {volume} {68}},\ \bibinfo
  {pages} {043814} (\bibinfo {year} {2003})}\BibitemShut {NoStop}%
\bibitem [{\citenamefont {Achilles}\ \emph {et~al.}(2004)\citenamefont
  {Achilles}, \citenamefont {Silberhorn}, \citenamefont {Sliwa}, \citenamefont
  {Banaszek}, \citenamefont {Walmsley}, \citenamefont {Fitch}, \citenamefont
  {Jacobs}, \citenamefont {Pittman},\ and\ \citenamefont
  {Franson}}]{achilles_photon-number-resolving_2004}%
  \BibitemOpen
  \bibfield  {author} {\bibinfo {author} {\bibfnamefont {D.}~\bibnamefont
  {Achilles}}, \bibinfo {author} {\bibfnamefont {C.}~\bibnamefont
  {Silberhorn}}, \bibinfo {author} {\bibfnamefont {C.}~\bibnamefont {Sliwa}},
  \bibinfo {author} {\bibfnamefont {K.}~\bibnamefont {Banaszek}}, \bibinfo
  {author} {\bibfnamefont {I.~A.}\ \bibnamefont {Walmsley}}, \bibinfo {author}
  {\bibfnamefont {M.~J.}\ \bibnamefont {Fitch}}, \bibinfo {author}
  {\bibfnamefont {B.~C.}\ \bibnamefont {Jacobs}}, \bibinfo {author}
  {\bibfnamefont {T.~B.}\ \bibnamefont {Pittman}}, \ and\ \bibinfo {author}
  {\bibfnamefont {J.~D.}\ \bibnamefont {Franson}},\ }\href {\doibase
  10.1080/09500340408235288} {\bibfield  {journal} {\bibinfo  {journal}
  {Journal of Modern Optics}\ }\textbf {\bibinfo {volume} {51}},\ \bibinfo
  {pages} {1499} (\bibinfo {year} {2004})}\BibitemShut {NoStop}%
\bibitem [{\citenamefont {Waks}\ \emph {et~al.}(2004)\citenamefont {Waks},
  \citenamefont {Diamanti}, \citenamefont {Sanders}, \citenamefont {Bartlett},\
  and\ \citenamefont {Yamamoto}}]{waks_direct_2004}%
  \BibitemOpen
  \bibfield  {author} {\bibinfo {author} {\bibfnamefont {E.}~\bibnamefont
  {Waks}}, \bibinfo {author} {\bibfnamefont {E.}~\bibnamefont {Diamanti}},
  \bibinfo {author} {\bibfnamefont {B.~C.}\ \bibnamefont {Sanders}}, \bibinfo
  {author} {\bibfnamefont {S.~D.}\ \bibnamefont {Bartlett}}, \ and\ \bibinfo
  {author} {\bibfnamefont {Y.}~\bibnamefont {Yamamoto}},\ }\href {\doibase
  10.1103/PhysRevLett.92.113602} {\bibfield  {journal} {\bibinfo  {journal}
  {Physical Review Letters}\ }\textbf {\bibinfo {volume} {92}},\ \bibinfo
  {pages} {113602} (\bibinfo {year} {2004})}\BibitemShut {NoStop}%
\bibitem [{\citenamefont {Jiang}\ \emph {et~al.}(2007)\citenamefont {Jiang},
  \citenamefont {Dauler},\ and\ \citenamefont
  {Chang}}]{jiang_photon-number-resolving_2007}%
  \BibitemOpen
  \bibfield  {author} {\bibinfo {author} {\bibfnamefont {L.~A.}\ \bibnamefont
  {Jiang}}, \bibinfo {author} {\bibfnamefont {E.~A.}\ \bibnamefont {Dauler}}, \
  and\ \bibinfo {author} {\bibfnamefont {J.~T.}\ \bibnamefont {Chang}},\ }\href
  {\doibase 10.1103/PhysRevA.75.062325} {\bibfield  {journal} {\bibinfo
  {journal} {Physical Review A}\ }\textbf {\bibinfo {volume} {75}},\ \bibinfo
  {pages} {062325} (\bibinfo {year} {2007})}\BibitemShut {NoStop}%
\bibitem [{\citenamefont {Heilmann}\ \emph {et~al.}(2016)\citenamefont
  {Heilmann}, \citenamefont {Sperling}, \citenamefont {Perez-Leija},
  \citenamefont {Gräfe}, \citenamefont {Heinrich}, \citenamefont {Nolte},
  \citenamefont {Vogel},\ and\ \citenamefont
  {Szameit}}]{heilmann_harnessing_2016}%
  \BibitemOpen
  \bibfield  {author} {\bibinfo {author} {\bibfnamefont {R.}~\bibnamefont
  {Heilmann}}, \bibinfo {author} {\bibfnamefont {J.}~\bibnamefont {Sperling}},
  \bibinfo {author} {\bibfnamefont {A.}~\bibnamefont {Perez-Leija}}, \bibinfo
  {author} {\bibfnamefont {M.}~\bibnamefont {Gräfe}}, \bibinfo {author}
  {\bibfnamefont {M.}~\bibnamefont {Heinrich}}, \bibinfo {author}
  {\bibfnamefont {S.}~\bibnamefont {Nolte}}, \bibinfo {author} {\bibfnamefont
  {W.}~\bibnamefont {Vogel}}, \ and\ \bibinfo {author} {\bibfnamefont
  {A.}~\bibnamefont {Szameit}},\ }\href {\doibase 10.1038/srep19489} {\bibfield
   {journal} {\bibinfo  {journal} {Scientific Reports}\ }\textbf {\bibinfo
  {volume} {6}},\ \bibinfo {pages} {19489} (\bibinfo {year}
  {2016})}\BibitemShut {NoStop}%
\bibitem [{\citenamefont {Sperling}\ \emph {et~al.}(2012)\citenamefont
  {Sperling}, \citenamefont {Vogel},\ and\ \citenamefont
  {Agarwal}}]{sperling_true_2012}%
  \BibitemOpen
  \bibfield  {author} {\bibinfo {author} {\bibfnamefont {J.}~\bibnamefont
  {Sperling}}, \bibinfo {author} {\bibfnamefont {W.}~\bibnamefont {Vogel}}, \
  and\ \bibinfo {author} {\bibfnamefont {G.~S.}\ \bibnamefont {Agarwal}},\
  }\href {\doibase 10.1103/PhysRevA.85.023820} {\bibfield  {journal} {\bibinfo
  {journal} {Physical Review A}\ }\textbf {\bibinfo {volume} {85}},\ \bibinfo
  {pages} {023820} (\bibinfo {year} {2012})}\BibitemShut {NoStop}%
\bibitem [{\citenamefont {Diels}\ and\ \citenamefont
  {Rudolph}(2006)}]{diels_ultrashort_2006}%
  \BibitemOpen
  \bibfield  {author} {\bibinfo {author} {\bibfnamefont {J.-C.}\ \bibnamefont
  {Diels}}\ and\ \bibinfo {author} {\bibfnamefont {W.}~\bibnamefont
  {Rudolph}},\ }\href@noop {} {\emph {\bibinfo {title} {Ultrashort {Laser}
  {Pulse} {Phenomena}, {Second} {Edition}}}},\ \bibinfo {edition} {2nd}\ ed.,\
  edited by\ \bibinfo {editor} {\bibfnamefont {P.~F.}\ \bibnamefont {Liao}}\
  and\ \bibinfo {editor} {\bibfnamefont {P.}~\bibnamefont {Kelley}}\ (\bibinfo
  {publisher} {Academic Press},\ \bibinfo {address} {Amsterdam ; Boston},\
  \bibinfo {year} {2006})\BibitemShut {NoStop}%
\bibitem [{\citenamefont {{Corning Optical
  Fibre}}(2007)}]{corning_optical_fibre_optical_2007}%
  \BibitemOpen
  \bibfield  {author} {\bibinfo {author} {\bibnamefont {{Corning Optical
  Fibre}}},\ }\href {www.princetel.com/datasheets/SMF28e.pdf} {\enquote
  {\bibinfo {title} {Optical {Specifications} of {SMF}28e fibres},}\ }
  (\bibinfo {year} {2007})\BibitemShut {NoStop}%
\bibitem [{\citenamefont {Annunziata}\ \emph {et~al.}(2010)\citenamefont
  {Annunziata}, \citenamefont {Quaranta}, \citenamefont {Santavicca},
  \citenamefont {Casaburi}, \citenamefont {Frunzio}, \citenamefont {Ejrnaes},
  \citenamefont {Rooks}, \citenamefont {Cristiano}, \citenamefont {Pagano},
  \citenamefont {Frydman},\ and\ \citenamefont
  {Prober}}]{annunziata_reset_2010}%
  \BibitemOpen
  \bibfield  {author} {\bibinfo {author} {\bibfnamefont {A.~J.}\ \bibnamefont
  {Annunziata}}, \bibinfo {author} {\bibfnamefont {O.}~\bibnamefont
  {Quaranta}}, \bibinfo {author} {\bibfnamefont {D.~F.}\ \bibnamefont
  {Santavicca}}, \bibinfo {author} {\bibfnamefont {A.}~\bibnamefont
  {Casaburi}}, \bibinfo {author} {\bibfnamefont {L.}~\bibnamefont {Frunzio}},
  \bibinfo {author} {\bibfnamefont {M.}~\bibnamefont {Ejrnaes}}, \bibinfo
  {author} {\bibfnamefont {M.~J.}\ \bibnamefont {Rooks}}, \bibinfo {author}
  {\bibfnamefont {R.}~\bibnamefont {Cristiano}}, \bibinfo {author}
  {\bibfnamefont {S.}~\bibnamefont {Pagano}}, \bibinfo {author} {\bibfnamefont
  {A.}~\bibnamefont {Frydman}}, \ and\ \bibinfo {author} {\bibfnamefont
  {D.~E.}\ \bibnamefont {Prober}},\ }\href {\doibase 10.1063/1.3498809}
  {\bibfield  {journal} {\bibinfo  {journal} {Journal of Applied Physics}\
  }\textbf {\bibinfo {volume} {108}},\ \bibinfo {pages} {084507} (\bibinfo
  {year} {2010})}\BibitemShut {NoStop}%
\bibitem [{\citenamefont {{Evanescent Optics
  Inc.}}(2016)}]{smolenaars_low_2016}%
  \BibitemOpen
  \bibfield  {author} {\bibinfo {author} {\bibnamefont {{Evanescent Optics
  Inc.}}},\ }\href@noop {} {\enquote {\bibinfo {title} {Low loss in
  fibre-integrated beam splitters},}\ }\bibinfo {howpublished} {private
  communication} (\bibinfo {year} {2016})\BibitemShut {NoStop}%
\bibitem [{\citenamefont {Bohmann}\ \emph {et~al.}(2016)\citenamefont
  {Bohmann}, \citenamefont {Kruse}, \citenamefont {Sperling}, \citenamefont
  {Silberhorn},\ and\ \citenamefont {Vogel}}]{bohmann_direct_2016}%
  \BibitemOpen
  \bibfield  {author} {\bibinfo {author} {\bibfnamefont {M.}~\bibnamefont
  {Bohmann}}, \bibinfo {author} {\bibfnamefont {R.}~\bibnamefont {Kruse}},
  \bibinfo {author} {\bibfnamefont {J.}~\bibnamefont {Sperling}}, \bibinfo
  {author} {\bibfnamefont {C.}~\bibnamefont {Silberhorn}}, \ and\ \bibinfo
  {author} {\bibfnamefont {W.}~\bibnamefont {Vogel}},\ }\href
  {http://arxiv.org/abs/1611.04779} {\bibfield  {journal} {\bibinfo  {journal}
  {arXiv:1611.04779 [quant-ph]}\ } (\bibinfo {year} {2016})},\ \bibinfo {note}
  {arXiv: 1611.04779}\BibitemShut {NoStop}%
\bibitem [{\citenamefont {Zambra}\ \emph {et~al.}(2005)\citenamefont {Zambra},
  \citenamefont {Andreoni}, \citenamefont {Bondani}, \citenamefont {Gramegna},
  \citenamefont {Genovese}, \citenamefont {Brida}, \citenamefont {Rossi},\ and\
  \citenamefont {Paris}}]{zambra_experimental_2005}%
  \BibitemOpen
  \bibfield  {author} {\bibinfo {author} {\bibfnamefont {G.}~\bibnamefont
  {Zambra}}, \bibinfo {author} {\bibfnamefont {A.}~\bibnamefont {Andreoni}},
  \bibinfo {author} {\bibfnamefont {M.}~\bibnamefont {Bondani}}, \bibinfo
  {author} {\bibfnamefont {M.}~\bibnamefont {Gramegna}}, \bibinfo {author}
  {\bibfnamefont {M.}~\bibnamefont {Genovese}}, \bibinfo {author}
  {\bibfnamefont {G.}~\bibnamefont {Brida}}, \bibinfo {author} {\bibfnamefont
  {A.}~\bibnamefont {Rossi}}, \ and\ \bibinfo {author} {\bibfnamefont
  {M.~G.~A.}\ \bibnamefont {Paris}},\ }\href {\doibase
  10.1103/PhysRevLett.95.063602} {\bibfield  {journal} {\bibinfo  {journal}
  {Physical Review Letters}\ }\textbf {\bibinfo {volume} {95}},\ \bibinfo
  {pages} {063602} (\bibinfo {year} {2005})}\BibitemShut {NoStop}%
\bibitem [{\citenamefont {Dodonov}\ \emph {et~al.}(2006)\citenamefont
  {Dodonov}, \citenamefont {Mizrahi},\ and\ \citenamefont
  {Dodonov}}]{dodonov_engineering_2006}%
  \BibitemOpen
  \bibfield  {author} {\bibinfo {author} {\bibfnamefont {A.~V.}\ \bibnamefont
  {Dodonov}}, \bibinfo {author} {\bibfnamefont {S.~S.}\ \bibnamefont
  {Mizrahi}}, \ and\ \bibinfo {author} {\bibfnamefont {V.~V.}\ \bibnamefont
  {Dodonov}},\ }\href {\doibase 10.1103/PhysRevA.74.033823} {\bibfield
  {journal} {\bibinfo  {journal} {Physical Review A}\ }\textbf {\bibinfo
  {volume} {74}},\ \bibinfo {pages} {033823} (\bibinfo {year}
  {2006})}\BibitemShut {NoStop}%
\bibitem [{\citenamefont {Allevi}\ \emph {et~al.}(2010)\citenamefont {Allevi},
  \citenamefont {Bondani},\ and\ \citenamefont
  {Andreoni}}]{allevi_photon-number_2010}%
  \BibitemOpen
  \bibfield  {author} {\bibinfo {author} {\bibfnamefont {A.}~\bibnamefont
  {Allevi}}, \bibinfo {author} {\bibfnamefont {M.}~\bibnamefont {Bondani}}, \
  and\ \bibinfo {author} {\bibfnamefont {A.}~\bibnamefont {Andreoni}},\ }\href
  {\doibase 10.1364/OL.35.001707} {\bibfield  {journal} {\bibinfo  {journal}
  {Optics Letters}\ }\textbf {\bibinfo {volume} {35}},\ \bibinfo {pages} {1707}
  (\bibinfo {year} {2010})}\BibitemShut {NoStop}%
\end{thebibliography}%

\end{document}